\documentclass[fleqn,usenatbib]{mnras}
\usepackage{graphicx}	% Including figure files
\usepackage{amsmath}	% Advanced maths commands

\usepackage{txfonts}
\usepackage[T1]{fontenc}
\usepackage{ae,aecompl}

\title[Radio Flares from Collisions of Neutron Stars with Interstellar Asteroids]{Radio Flares from Collisions of Neutron Stars with Interstellar Asteroids}

\author[A. Siraj and A. Loeb]{
Amir Siraj,$^{1}$\thanks{amir.siraj@cfa.harvard.edu}
Abraham Loeb,$^{1}$\thanks{aloeb@cfa.harvard.edu}
\\
$^{1}$Department of Astronomy, Harvard University, 60 Garden Street, Cambridge, MA 02138, USA\\
}

\date{Accepted XXX. Received YYY; in original form ZZZ}

\pubyear{2019}

\begin{document}
\label{firstpage}
\pagerange{\pageref{firstpage}--\pageref{lastpage}}
\maketitle

\begin{abstract}
We propose that collisions between neutron stars and interstellar asteroids, such as `Oumuamua, could power observable radio flares in the Milky Way galaxy. We find the rate of such events at $\sim 1 \mathrm{\; Jy}$ to be $\sim 10 \mathrm{\; day^{-1}}$.
\end{abstract}


%%%%%%%%%%%%%%%%%%%%%%%%%%%%%%%%%%%%%%%%%%%%%%%%%%

%%%%%%%%%%%%%%%%% BODY OF PAPER %%%%%%%%%%%%%%%%%%

\section{Introduction}
Fast Radio Bursts (FRBs) are observed to have millisecond duration at a frequency of $\sim 1$ GHz \citep{Katz2019}. A multitude of hypotheses exist to explain FRBs, many involving neutron stars (NSs) \citep{Platts2019}. \cite{Geng2015} proposed that FRBs may be powered by collisions between NSs and asteroids/comets, and \cite{Dai2016} studied the acceleration and radiation mechanisms of ultra-relativistic electrons in such collisions while proposing that repeating FRBs could be explained by NSs traveling through asteroid belts. The \cite{Geng2015} hypothesis lacks a clear source of asteroids to power non-repeating FRBs.

Rotating Radio Transients (RRATs) have durations and frequencies similar to FRBs, but they originate in Milky Way galaxy and their luminosities are a billion times fainter \citep{McLaughlin2006}.

`Oumuamua and CNEOS 2014-01-08 represent the first interstellar asteroids (ISAs) larger than dust discovered in the Solar System \citep{Meech2017, Micheli2018, Siraj2019a}, serving as calibrations for the ISA size distribution \citep{Siraj2019b}.

In this \textit{Letter}, we explore the possibility that ISAs could power a subclass of RRATs which do not repeat over long timescales. The outline of the paper is as follows. In Section~\ref{sec:emission} we summarize the emission of collisions of NSs with ISAs. In Section~\ref{sec:method} we describe the method used to derive the observable event rate, and in Section~\ref{sec:rate} we report the result. Finally, Section~\ref{sec:discussion} summarizes our main conclusions.

\section{Emission Mechanism}
\label{sec:emission}

\cite{Dai2016} theorize that during each NS-asteroid impact, electrons are torn off the tidally-disrupted asteroid, accelerated to ultra-relativistic energies instantaneously as they travel along magnetic field lines and emit coherent curvature radiation. Assuming an NS radius of $\sim 12.5$ km \citep{Abbott2018, Greif2019}, the luminosity per unit frequency at $\nu \sim 1 \mathrm{\; GHz}$ is,

\begin{equation}
\begin{aligned}
    L_{\nu} \approx \; 3.1 \times 10^{21} 
    & \left(\frac{M_{NS}}{1.4M_{\odot}}\right)^{19/12} \left(\frac{\mu_{NS}}{10^{30} \mathrm{\; G\;cm^{-2}}}\right)^{3/2} \\
    &\left(\frac{\kappa}{0.13} \right)^{-1} \left(\frac{s}{10^{10}\mathrm{\; dyn\;cm^{-2}}} \right)^{2/3}
    \\
    &  \left(\frac{\rho}{8\mathrm{\; g\;cm^{-3}}} \right)^{-14/9} \left(\frac{r}{\mathrm{1\;m}}\right)^{8/3}
    \mathrm{\;erg\; s^{-1} \; Hz^{-1},} \\
\end{aligned}
\end{equation}
where $M_{NS}$ is the NS mass, $\mu_{NS}$ is the NS magnetic dipole moment, $\kappa$ is a constant related to the tensile and compressive strengths of the (iron-rich) asteroid, $s$ is the tensile strength of the asteroid, $\rho$ is the mass density of the asteroid, and $r$ is the radius of the asteroid. \cite{Dai2016} also show the duration  of the emission to be of order a millisecond due to the time difference between the leading and lagging fragments of the tidally disrupted asteroid impacting the NS.

\section{Method}
\label{sec:method}

\begin{figure}
  \centering
  \includegraphics[width=.9\linewidth]{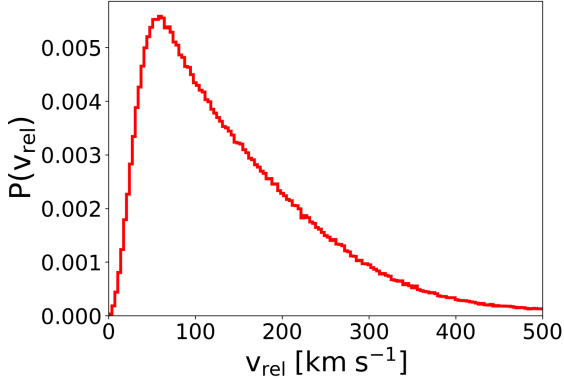}
    \caption{Calculated probability function of the relative speeds between NSs and ISAs.}
    \label{fig:1}
\end{figure}

We adopt the three--dimensional velocity dispersions for stars in the thin disk of the Milky Way as a proxy for the kinematics of ISAs, each corresponding to the standard deviation of a Gaussian distribution about the local standard of rest (LSR): $\sigma_x = 35 \mathrm{\;km\;s^{-1}}$, $\sigma_y = 25 \mathrm{\;km\;s^{-1}}$, $\sigma_z = 25 \mathrm{\;km\;s^{-1}}$ \citep{Bland-Hawthorn2016}. We take the velocity distribution of NS (relative to the LSR) to be a two-component Gaussian described by the following probability function \citep{Faucher-Giguere2005}:

\begin{equation}
    P(v_{NS}) \approx \frac{w_1}{\sqrt{2\pi}\sigma_1} \exp\left({\frac{v_{NS}^2}{\sigma_1^2}}\right) + \frac{1 - w_1}{\sqrt{2\pi}\sigma_2}\exp\left( {\frac{v_{NS}^2}{\sigma_2^2}}\right)\; \; ,
\end{equation}
where $w_1 = 0.90$, $\sigma_1 = 160 \mathrm{\;km\;s^{-1}}$, and $\sigma_2 = 780 \mathrm{\;km\;s^{-1}}$.

We use a Monte Carlo method to determine the characteristic relative speed of NS-ISA collisions, $\tilde{v}_{rel}$. First, we draw randomly from the Gaussian distributions described by the velocity ellipsoid for ISAs. We then draw from the two-component Gaussian distribution describing $P(v_{NS})$ and choose a random direction on the unit sphere to determine the components of $v_{NS}$. The magnitude of the difference of the two velocity vectors, ${v}_{rel}$, is then computed. The results of the Monte Carlo method are shown in Fig.~\ref{fig:1}. The median relative speed is, $\tilde{v}_{rel} = 130 \mathrm{\; km \; s^{-1}}$.

We use the expression derived by \cite{Dai2016}, based on \cite{Safronov1972}, to derive the NS-ISA impact cross section, including gravitational focusing, of $\sigma_a \approx 1.7 \times 10^{19} \left( \frac{v_{rel}}{100 \; \mathrm{km\;s^{-1}}}\right)^{-2} \mathrm{\;cm^{2}}$.

Assuming that the distributions of ISAs and NSs follow the distribution of stars \citep{Faucher-Giguere2005}, we define $\zeta_\mathrm{ISA} \equiv n_\mathrm{ISA} / n_{\star}$ and $\zeta_\mathrm{NS} \equiv n_\mathrm{NS} / n_{\star}$.

The cumulative Earth impact rate for an ISA of radius $r$ is estimated to be $2 \times 10^{-4} \left(r/\mathrm{1 \:m} \right)^{-3.4}$ \citep{Siraj2019b}. Assuming that $\sim 5\%$ of all asteroids are composed primarily of iron \citep{Burbine2002}, we find the number density of ISAs of radius $\geq r$ to be related to the number density of stars by a factor of, 

\begin{equation}
\zeta_{ISA} \sim 2.5 \times 10^{18} \left(\frac{1 \; \mathrm{pc}^{-3}}{n_{\star,\; \odot}} \right) \left(\frac{r}{\mathrm{1 \:m}} \right)^{-3.4} \; \; .
\end{equation}
where $n_{\star,\; \odot}$ is the number density of stars in the solar neighborhood.

The minimum luminosity at a frequency of $\nu \sim 1 \mathrm{\;GHz}$ for a source at a distance $d$ to be visible with a detector of flux threshold $f$ is,

\begin{equation}
    L_\nu = 4 \pi \times 10^{14} \left(\frac{f}{\mathrm{1\;Jy}} \right) \left(\frac{d}{\mathrm{1\;pc}} \right)^2 \; \mathrm{erg \; s^{-1} \; Hz^{-1}} \; \; ,
\end{equation}
yielding the minimum ISA radius that produces a visible flare,

\begin{equation}
    r \sim 0.71 \left(\frac{d}{\mathrm{kpc}} \right)^{3/4} \left(\frac{f}{\mathrm{Jy}} \right)^{3/8} \; \mathrm{m} \; \;,
\end{equation}
thereby allowing us to express $\zeta_{ISA}$ in terms of $d$ and $f$ as,

\begin{equation}
    \zeta_{ISA} \sim 8.1 \times 10^{18} \left(\frac{1\; \mathrm{pc}^{-3}}{n_{\star,\; \odot}\;} \right) \left(\frac{d}{\mathrm{1\;kpc}} \right)^{-2.55} \left(\frac{f}{\mathrm{1\;Jy}} \right)^{-1.28} \; \; .
\end{equation}

We model the Milky Way galaxy as a disk with a radial scale length $R_d \sim 3 \; \mathrm{kpc}$ and vertical scale height $h \sim 0.1 \; R$ as a function of radial distance $R$ and vertical distance $z$, 

\begin{equation}
    n_{\star} \propto \exp{\left(-\frac{R}{R_d} \right)} \exp{\left(-\frac{\lvert z \rvert}{h} \right)} \; \;.
\end{equation}

\begin{figure}
  \centering
  \includegraphics[width=.9\linewidth]{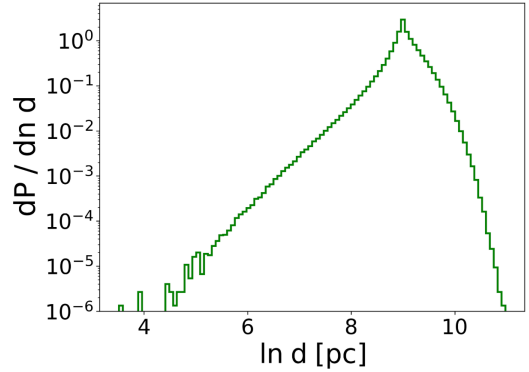}
    \caption{Normalized probability function of the distance of NS from the Earth.}
    \label{fig:2}
\end{figure}

Given $\zeta_{NS} = 1.7 \times 10^{-3}$, we compute randomly generated positions of $10^{8}$ NSs \citep{Sartore2010} in the Galaxy (following the density of stars), and subsequently find the distance between each one and the Earth. The associated probability distribution is shown in Fig.~\ref{fig:2}; the transition in slope around $d \sim h$ is caused by the change from a 3D to a 2D distribution of sources. We then find the rate at which each NS produces flares at or above the limiting flux $f$ as measured from Earth to be,

\begin{equation}
    \dot{N}_{flare,\;NS} = \zeta_\mathrm{ISA} n_{\star}\; \sigma_a {v}_{rel} \; \; \mathrm{,}
\end{equation}
where $v_{rel}$ is drawn from the aforementioned Monte Carlo method. Finally, we sum the rates for each individual NSs to find the total rate of visible NS-ISA flares. The minimum asteroid radius considered, $r_{min}$, is given approximately by the minimum size at which an asteroid is tidally disrupted before it reaches its melting point. \cite{Cordes2008} and \cite{Geng2015} conclude that $r_{min} \sim 1 \; \mathrm{m}$, as below this size, an iron asteroid will melt before it is tidally disrupted.

\section{Rate}
\label{sec:rate}

We find the all-sky rate of observable NS-ISA flares to be described as the following fitting function,

\begin{equation}
\label{eq:fitting}
\dot{N} \sim
\begin{cases}
    \left( 8.2 - 0.8\left( r_{min}/1\;\mathrm{m} \right) \right) \left( f /\mathrm{1 \;Jy} \right)^{-1.28} \mathrm{\; day^{-1}} & \text{if $r_{min} \leq 3.4 \mathrm{\; m}$} \\
    \left(370 \left( r_{min}/1\;\mathrm{m} \right)^{-3.4} \right) \left( f /\mathrm{1 \;Jy} \right)^{-1.28} \mathrm{\; day^{-1}} & \text{if $r_{min} > 3.4 \mathrm{\; m}$} \; \;.
\end{cases}
\end{equation}

\begin{figure}
  \centering
  \includegraphics[width=.9\linewidth]{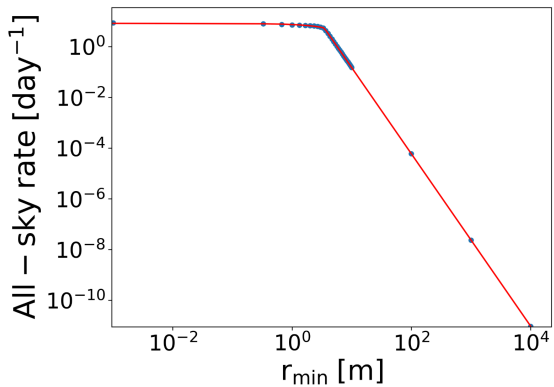}
    \caption{Rate of $\sim 1$ Jy radio flares at $\nu \sim$ 1 GHz from NS-ISA collisions, as a function of the minimum ISA radius, $r_{min}$. The red line shows the piece-wise fitting function in Equation~(\ref{eq:fitting}).}
    \label{fig:3}
\end{figure}

Fig.~\ref{fig:3} shows the rate as a function of $r_{min}$ along with an associated fitting function.

\section{Discussion}
\label{sec:discussion}

We have shown that NS-ISA collisions could reliably power observable, non-repeating, millisecond-duration $\sim 1$ GHz radio flares in the Milky Way galaxy. We would not expect to detect any X-ray emission from such events, given the expression for X-ray flux in \cite{Dai2016} for $r_{min} \gtrsim 1 \; \mathrm{m}$. We do not expect such events to constitute a significant fraction of FRBs due to the low abundance of sufficiently large asteroids to produce observable flares at cosmological distances. Our rate is also too small to explain the FRB rate from the Milky Way galaxy alone.

The abundance of single-pulse RRATs is still poorly constrained, so it is difficult to compare our estimated rate with the total estimated rate (Agarwal, McLaughlin \& Lorimer, private communication).

Most of the asteroids around the original progenitor star that exploded were lost because the star lost most of its mass, and so their energy relative to the NS remnant became positive, but new asteroids (as well as planets\footnote{https://en.wikipedia.org/wiki/Pulsar\_planet}) may form out of the post supernova debris.

NS-ISA collisions represent a new class of transients that could reveal the distributions and abundances of both NSs and ISAs, serving as an important calibration for both populations.

%\vspace{0.1in} 
%\pagebreak
%\newpage
\section*{Acknowledgements}
%\vspace{0.1in} 
This work was supported in part by a grant from the Breakthrough Prize Foundation. %\newline \newline

\bsp	% typesetting comment
\label{lastpage}
\end{document}